\documentclass[conference]{IEEEtran}
\IEEEoverridecommandlockouts
\usepackage{cite}
\usepackage{amsmath,amssymb,amsfonts}
\usepackage{algorithmic}
\usepackage{graphicx}
\usepackage{textcomp}
\usepackage{xcolor}
\usepackage{xspace}
\usepackage{url}
\def\BibTeX{{\rm B\kern-.05em{\sc i\kern-.025em b}\kern-.08em
    T\kern-.1667em\lower.7ex\hbox{E}\kern-.125emX}}

\newcommand{\tool}{\textsc{Suggestion Bot}\xspace}

\begin{document}

\title{Suggestion Bot: Analyzing the Impact of Automated Suggested Changes on Code Reviews
}

\author{\IEEEauthorblockN{Nivishree Palvannan}
\IEEEauthorblockA{\textit{Department of Computer Science} \\
\textit{Virginia Tech}\\
Virginia, USA \\
nivipal@vt.edu}
\and
\IEEEauthorblockN{Chris Brown}
\IEEEauthorblockA{\textit{Department of Computer Science} \\
\textit{Virginia Tech}\\
Virginia, USA \\
dcbrown@vt.edu}
}

\maketitle

\begin{abstract}
Peer code reviews are crucial for maintaining the quality of the code in software repositories. Developers have introduced a number of software bots to help with the code review process. Despite the benefits of automating code review tasks, many developers face challenges interacting with these bots due to non-comprehensive feedback and disruptive notifications.  In this paper, we analyze how incorporating a bot in software development cycle will decrease turnaround time of pull request. We created a bot called ``\tool''  to automatically review the code base using GitHub's suggested changes functionality in order to solve this issue. A preliminary comparative empirical investigation between the utilization of this bot and manual review procedures was also conducted in this study. We evaluate \tool concerning its impact on review time and also analyze whether the comments given by the bot are clear and useful for users. Our results provide implications for the design of future systems and improving human-bot interactions for code review.
\end{abstract}

\begin{IEEEkeywords}
Pull Requests, Peer Code Reviews, Code Review Bot, Suggested changes
\end{IEEEkeywords}

\section{Introduction}
Software development on GitHub is pull-based~\cite{gousious2014pullbased}, which allows branching and isolated development for individuals in a distributed software engineering team to make changes to a central repository. Millions of both open- and closed-source projects utilize hosting sites like GitHub\footnote{\url{https://github.com}} with pull-based development features~\cite{Octoverse}. In software development processes on GitHub, pull requests are a way to collaborate and inspect changes. The inspection of pull requests (PRs) typically consists of a \textit{peer code review} of the supplied commits, or code changes from a contributor reviewed by a project maintainer with a discussion of the changes made in the pull request. The reviewer will examine the code modifications, review them, and then push them to the master branch. The process for a pull request approval in GitHub will involve getting the project maintainer(s) or peer workers to review your work; after which they will provide comments or, if your pull request is approved, will merge your changes directly into the main repository. Pull-requests as implemented by GitHub are a model for collaborating on distributed software  development. 
 
 In spite of having various advantages for improving code quality~\cite{mcintosh2016empirical} and teamwork~\cite{trytten2005design}, there are various challenges with peer code reviews. For example, prior work suggests modern code review practices are time-consuming and slow~\cite{czerwonka2015slows}, require in-depth understanding of code bases~\cite{bacchelli2013expectations}, and incorporate bias based on gender~\cite{nasif2019gender} and race~\cite{nadri2022race}. Another key disadvantage of peer code reviews is the large burden placed on code reviewers~\cite{gousios2015integrator}. Many pull requests in GitHub repositories are stagnated due to lack of code reviews~\cite{yu2015wait}. This increases the likelihood that PRs will be rejected or ignored, ultimately discouraging developers from making future contributions to projects~\cite{legay2019future}. Moreover, complaints from developers on code review processes include untimely and unuseful or incomprehensible feedback from reviewers~\cite{macleod2018trenches}.

Projects have adopted bots to automate peer code review tasks, thereby assisting PR integrators and contributors in their work. Bots such as Review Bot have been shown to improve code quality while reducing reviewer effort at VMWare~\cite{balachandran2013reviewbot}. Further, Wessel et al. show that code review bots and automated tools are useful for reviewing processes and increase the number of pull requests merged~\cite{wessel2020effects}. However, prior work also suggests developers find bots challenging to work with in software development contexts. For instance, research shows software engineers report that bots have non-comprehensive feedback~\cite{wessel2018power} and interrupt development workflows~\cite{SorryToBotherYou}. These poor human-bot interactions lead to frustration and distractions for software developers~\cite{wessel2021disturb}. 
    
To overcome these challenges with code review bots, we developed a bot called \tool. This system aims to reduce the noise from automated bots in pull request reviews by taking advantage of the GitHub suggested changes feature to provide concise feedback to developers and reduce interruptions in code review processes. We conducted an experiment to investigate on the impact of using \tool for pull request reviewing. Our study explores our following research questions.

\textbf{\textit{RQ1}} How quickly are pull requests are reviewed using \tool?

\textbf{\textit{RO2}}  How  useful are recommendations from suggested changes using a bot while reviewing pull requests?


To analyze the usage and impact of our bot, we conducted a preliminary user study with 5 participants who have some prior experience with pull request reviews on GitHub. Study tasks included manual code review and code review using \tool to compare and contrast the advantages of incorporating the bot into the development cycle. From the results of the study, we provide initial evidence suggesting \tool provides clear and understandable feedback and decreases the turnaround time of PRs in code review processes.



\begin{figure}[t!]
    \centering
    \includegraphics[width=0.4\textwidth]{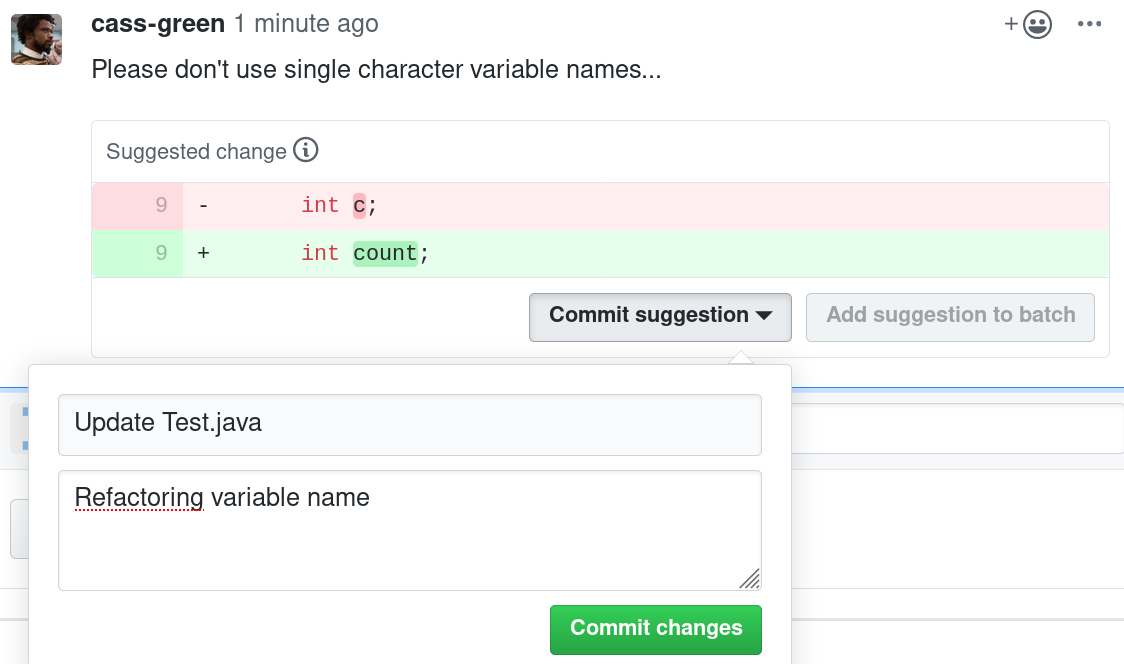}
    \caption{GitHub suggested changes example}
    \label{fig:change}
\end{figure}

\section{Background}
 	\subsection{GitHub Pull Requests and Peer Code Reviews} \label{se:another_section}
Pull requests are the primary feature for pull-based development on GitHub~\cite{gousious2014pullbased}.\footnote{\url{https://docs.github.com/en/pull-requests/collaborating-with-pull-requests/proposing-changes-to-your-work-with-pull-requests/about-pull-requests}} PRs rely on peer code reviews, a manual inspection of source code by project maintainers, to integrate code into the main branch. The authors of PRs can then apply or reject the code changes suggested by collaborators and reviewers through  comments as part of the code review process. In the past, code changes were reviewed through inspection where there were formal in-person meetings~\cite{fagan1976inspection}. However, current code review processes through PRs are asynchronous and support geographically distributed reviewers~\cite{ebert2001surviving}. 

\subsection{GitHub Suggested Changes}

In 2018, GitHub introduced \textit{suggested changes}, a new feature to provide feedback on pull requests.\footnote{\url{https://github.blog/changelog/2018-10-16-suggested-changes/}} After a PR is submitted to the repository, suggested changes allow code reviewers to make recommendations to contributors by suggesting specific lines of code on the PR. This feature also provides functionality for contributors to automatically apply, reject, or edit changes suggested by reviewers. An example is presented in Figure~\ref{fig:change}. GitHub users were ``quick to adopt suggested changes'' within a few weeks of its release, with development teams frequently utilizing them for code reviews and integrating this feature into their code review processes.\footnote{\url{https://github.blog/2018-11-01-suggested-changes-update/}} Prior work empirically investigated the suggested changes feature, and found this tool is useful for making recommendations to improve the quality of code during PR reviews because of its concise communication and effortless integration into existing workflows~\cite{SuggestedChanges}.

\subsection{Code Review Bots}

Researchers have implemented a wide variety of automated bots to support pull request review tasks. Bots such as Dependabot\footnote{\url{https://docs.github.com/en/code-security/dependabot/working-with-dependabot}} and Codecov\footnote{\url{https://docs.codecov.com/docs/team-bot}} provide useful information to reviewers on outdated package dependencies and code testing coverage. Review Bot~\cite{balachandran2013reviewbot} consolidates the output from various static analysis tools, which has been shown to reduce reviewer effort in reviews~\cite{singh2017evaluating}. Other bots, such as RevFinder~\cite{thongtanunam2015should}, automatically recommend reviewers for pull requests to prevent stagnation. In general, research suggests bots can enhance code review processes~\cite{wessel2020effects}. We aim to build upon this work by introducing a novel bot to support code reviews by providing concise feedback and minimal interruptions to workflows.

   
\section{Suggestion Bot}
\begin{figure*}[htp]
    \centering
    \includegraphics[width=0.6\textwidth]{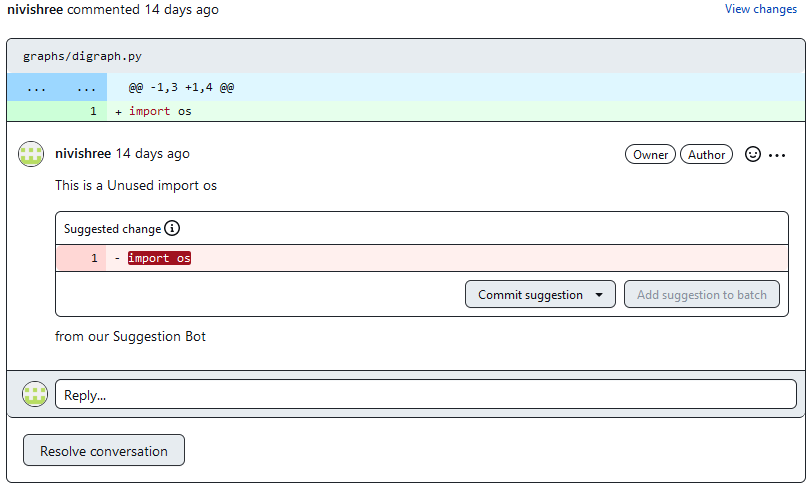}
    \caption{Recommendation from \tool on a pull request}
    \label{fig:galaxy}
\end{figure*}

To improve human-bot interactions during pull request reviews, we created a bot called \tool. The goal of this bot is to examine pull requests and provide timely, concise, and non-interruptive feedback to developers. To do this, we leverage the GitHub suggested changes feature as the primary feedback mechanism for \tool which provides clear feedback to users without interrupting existing workflows~\cite{SuggestedChanges}. \tool is able to analyze open pull requests on public repositories. The bot works by running static analysis tools on the modified version of contributors' code fetched using the GitHub API, and then generates recommendations for improvements to change the code by providing a suggested change automatically on the PR based on the static analysis tool output. An example suggestion from \tool is presented in Figure~\ref{fig:galaxy}.

For our initial implementation of \tool, we integrated Black\footnote{\url{https://black.readthedocs.io/en/stable/}} into the workflow of our bot. Black is a popular static code analysis tool for Python that is is PEP 8 compliant and can be used for static error identification. It is open-sourced and available in free versions for continuous inspection of code quality and style during peer code reviews. Our preliminary implementation of \tool involves code review analysis of Python code and automated suggested changes featuring formatted output from Black. We evaluate \tool by analyzing the effectiveness and advantages of this bot in comparison with manual effort during pull request code reviews.

\section{Study Methodology}

We devised a preliminary experiment to understand the effects of \tool on peer code review tasks.

\subsection{Participants}

Our preliminary evaluation consists of a user study where participants completed code review tasks with and without using \tool. We first asked participants to complete a demographic questionnaire. Using this initial survey, we hoped to gauge participants' knowledge of and experience with GitHub and reviewing pull requests. The participants were required to have some background in software engineering as well as a working knowledge of GitHub and reviewing pull requests to participate. Overall, we had seven participants. All subjects had some prior experience with software development and GitHub, however most participants were unfamiliar with the GitHub suggested changes feature. Participants were a mix of professional developers and students with prior industry experience, averaging about two years of professional software engineering work. In addition, participants reported using GitHub at least ``Most of the Time'' or ``Sometimes''.

\subsection{Study Tasks}

After the initial questionnaire, participants were asked to manually review of a pull request created by the authors for study purposes. The mock PR was created on an existing and popular public Python repository on GitHub. When customizing the pull request, we aimed to ensure the study environment was relevant to real-world software and also incorporated errors that reviewers and \tool would be able to detect. We aimed to design our task to reflect authentic code review tasks in a development environment and be reviewable in a limited amount of time for the user study (approx. 30 minutes). Participants were asked to think-aloud and make observations about the code and point out any potential errors in the Python code. After the manual review, participants were asked to review the same set of code using \tool. Participants were similarly asked to note any observations and differences between these two processes. To compare \tool with manual inspection, we observed the amount of time for pull requests to be reviewed both with and without the bot to understand the effects of \tool when performing code reviews. All study sessions were recorded for further retrospective analysis of the study tasks.

We concluded the study session with a post survey to gain user feedback on our bot and participants' experiences using it for code reviews. The survey used rating scale questions for participants to rank responses from 0 to 100. We were also interested in whether participants would be willing to adopt this tool and the clarity of feedback. This method gave us a more thorough and nuanced grasp of the numerous approaches to enhance the code review process and to enhance \tool in the future. For our study, we were specifically interested in the impact of our bot on pull request review times and the usefulness of feedback from \tool.

\subsubsection{Time}

Many pull requests experience delays in review due to variety of reasons~\cite{yu2015wait}. For instance, in the study pre-questionnaire participants reported experiencing delays in pull request reviews due to a variety of issues such as other work tasks and meetings. Subjects also reported PR reviews consume a lot of time due to the need to read the code, understand functionality, inspect for design issues and to assess the code quality, and test the new implementation. However, participants noted their ideal turnaround time for pull requests would be the same day or within one week. To measure the impact of \tool on time, we observed how long it took participants to review pull requests in the study tasks manually in comparison to reviewing them with \tool and asked them to rank our system based on its ability to decrease PR turnaround time. We hypothesize that peer code reviews with \tool will take less time compared to manual inspection of pull requests.

\subsubsection{Usefulness}

Pull requests are an intermediate step for merging code into source code repositories~\cite{gousious2014pullbased}. It is a mechanism where the developer will notify team members that a new contribution is ready to be merged to the main branch, along with serving as a discussion forum for developers to review the code changes. However, this feedback is not always useful. For example, in our pre-questionnaire most participants reported receiving ``Somewhat understandable'' comments on their own pull requests. Additionally, bots can provide ineffective feedback on PRs~\cite{wessel2018power}. To measure the usefulness of feedback from \tool, we debriefed participants after the study tasks with a post-survey to provide additional insight into their experience using our bot to review a pull request. We speculate participants will find \tool useful due to its concise and actionable feedback as well as its ability to seamlessly fit into code review processes.


\section{Results}


\subsection{RQ1: Time}

\begin{figure}[htp]
   \centering
   \includegraphics[width=0.45\textwidth]{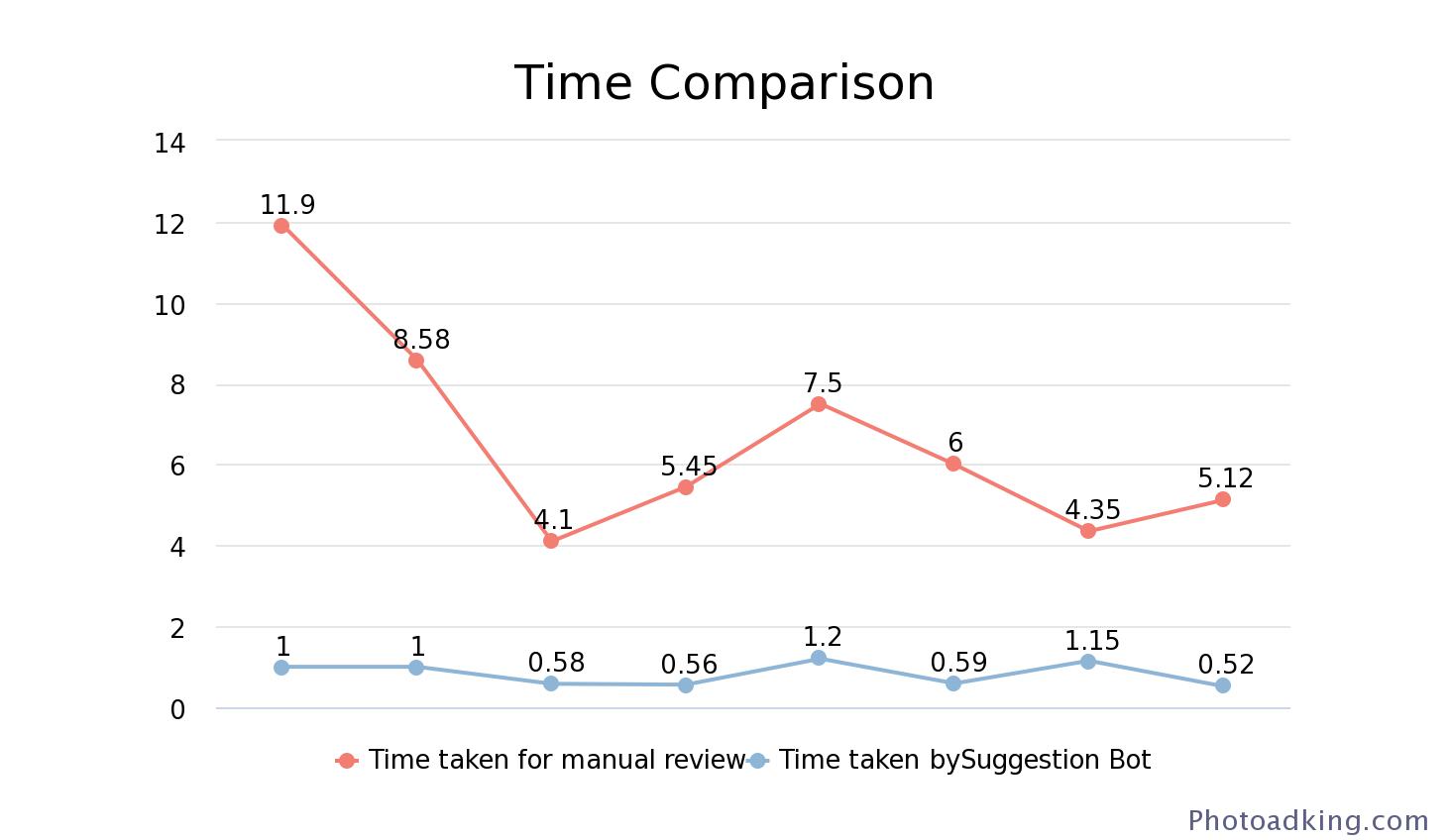}
   \caption{Time Comparison between manual code review and \tool}
   \label{fig:time}
\end{figure}
One important conclusion from the study is that manual reviewing takes more time to review and make suggestions in comparison to \tool. Participants averaged over seven minutes to manually inspect the pull request, while \tool itself averaged approximately 50 seconds to run and make comments on PRs. Using a t-test to compare the average review time in each setting, we found these results are statistically significant ($t = 5.67406, p-value = 0.0001$). Most participants spent time studying the code before providing review comments. Some participants mentioned they normally take their time to acquire and understand code that is novel or whose design is unknown before giving comments. We also found particular ranked \tool with a 97.4 on it's ability to reduce pull request review times. In addition to speeding up PR turnaround time, the tool also provided comments needed to improve the code quality and standards. Some participants completed the code review in two to three minutes, however there were some undetectable problems, such as white space issues and improper code styling. As opposed to this, employing \tool will find problems and make suggestions for improvements.

\subsection{RQ2: Usefulness}

To gain insight into the usefulness of \tool, we surveyed participants after completing the study tasks. With respect to the usefulness of \tool for reviewing pull requests, participants rated the bot with a 95, indicating they found it very useful for completing code review tasks. Further, \tool was ranked 86.4 for whether subjects believed it would be adaptable for new projects and 95.7 for whether or not respondents would suggest this tool to coworkers. These results suggest participants found \tool usable and would be willing to adopt this system for their own peer code review processes. Finally, concerning the feedback from \tool, participants ranked our system with a 92.1 for clarity and a 93 for general perceptions of the comments. We found participants appreciated the feedback comments from \tool using the suggested changes feature on GitHub. These results substantiate prior work, which shows the suggested changes feature is useful for code reviews and provides clear feedback to developers on pull requests.

\section{Discussion}

Our results suggest that bots are effective for influencing pull request reviews on GitHub, and participants found value in using \tool to support code review tasks. \tool was able to reduce the review time for PR reviews compared to manual inspection. However, as the majority of bots improve efficiency for completing manual tasks, we were also interested in the usability of \tool. All of our participants found our bot useful for code reviews and providing feedback to developers by providing high ratings on the usability of \tool---even expressing interest in adopting the tool for their own projects and recommending to colleagues. This points to a
need for design improvements for code review bots
to make better recommendations with clear feedback and reduced noise. Based on our experience implementing and evaluating \tool, we encourage researchers and toolsmiths look beyond improving code review tasks with automation and also consider implementing concise and actionable feedback that can be easily integrated into project workflows when developing future bots to support peer code reviews.

\section{Summary and Future Work}

Automated bots have been implemented to automate code review tasks and reduce developer effort during reviews. However, software bots often generate poor interactions with humans due to incomprehensible feedback and disruption of workflows. In this work, we introduce \tool as a novel system that utilizes GitHub suggested changes to provide feedback to submitted code. We analyzed the relationship between pull request code review processes with \tool and manual inspection, and found that our bot reduced the turnaround time of pull requests and provided concise and useful feedback to users. This paper makes a step towards improving the design of code review bots by providing actionable feedback and minimizing interruptions to development processes. Our ongoing and future work consists of enhancing \tool to work with other development tools for different programming languages beyond using Black to detect Pylint issues. Additionally, we plan to further evaluate the bot with more participants and compare with other code review bots to analyze the effects of \tool on pull request review processes.


\bibliography{main}
\bibliographystyle{abbrv}


\end{document}